\documentclass[12pt]{article}
\usepackage{amsxtra,amssymb,amsthm,amsmath,latexsym}

\theoremstyle{plain}

\newtheorem{theorem}{Theorem}[section]

\newtheorem{remark}{Remark}

\def\R{{\mathbb R}}

\def\oH{\buildrel\circ\over H}
\def\oH1{\buildrel\circ\over H\kern-.02in{}^1}
\def\qed{{\hfill $\Box$}}

\def\supp{\hbox{\,supp\,}}
\def\const{\hbox{\,const\,}}

\def\bysame{\rule{.5in}{.005in},\ }
\def\Im{\hbox{\,Im\,}}
\def\varep{\varepsilon}

\begin{document}


\title{
Justification of the limiting absorption principle in $\R^2$
   \thanks{key words:  limiting absorption principle, spectral
theory, elliptic equations }
   \thanks{Math subject classification: primary 35J15\quad secondary
35P25}
}

\author{
A.G. Ramm\\
 Mathematics Department, Kansas State University, \\
 Manhattan, KS 66506-2602, USA\\
ramm@math.ksu.edu\\
}

\date{}

\maketitle\thispagestyle{empty}

\begin{abstract}
Limiting absorption principle is justified for
second-order selfadjoint elliptic operators in
two-dimensional spaces.
\end{abstract}


\section{Introduction.
Statement of the problem.
Formulation of the results.}

In this paper we prove limiting absorption principle in 
two-dimensional space for second-order 
self-adjoint elliptic operators. This principle
has been discussed in the literature extensively
([1],[2], [4]-[8]). Our approach is similar to the one
developed in [4]-[6].

Let
\begin{equation}
  Lu:=-\nabla\cdot[a(x)\nabla u]=f(x)\hbox{\ \ in\ \ }\R^2.
  \tag{1.1}\end{equation}

Assume that
\begin{equation}
  a=a_{ij}=a_{ji}=\overline{a_{ij}},\quad 1\leq i,\ j\leq 2,
  \tag{1.2}\end{equation}
is a symmetric real-valued matrix,
the overbar stands for complex conjugate,
\begin{equation}
 a_0t_j\overline{t_j}\leq a_{ij}t_i\overline{t_j}\leq a_1 t_j\overline{t_j},
  \quad a_1>a_0=\const>0,
  \tag{1.3}\end{equation}
over the repeated indices summation is understood,
$a_1$ is a positive constant,
$a_{ij}(x)\in C^{0,1}(\R^2)$, and
\begin{equation}
  a_{ij}=\delta_{ij}\hbox{\quad for\quad} |x|>R,
  \tag{1.4}\end{equation}
where $R>0$ is an arbitrary large fixed number.

Let us assume that
\begin{equation}
  \supp f\subset B_R:=\{ x:|x|\leq R\},
  \quad f\in L^2(B_R).
  \tag{1.5}\end{equation}
We have chosen $f\in L^2$ in order to use
a priori estimates in the Sobolev spaces $H^2:=W^{2,2}$ for the
 selfadjoint realization of the operator $L$
in $L^2(\R^2)$.

By $B_R':=\R^2\backslash B_R$ the exterior domain will be
denoted. One could take in (1.5) $f\in L^p(B_R)$ with any $p>1$.
Then the solution to (1.1) $u\in W^{2,p}_{loc}(\R^2)$, where
$W^{\ell,p}$ is the Sobolev space, \cite{G},
and $C(\R^2)$ is the space of continuous functions.

We are looking for the solution to (1.1) which vanishes at
infinity in the sense:
\begin{equation}
  |u|=O\left(\frac{1}{|x|}\right), \quad
  |\nabla u|=O\left( \frac{1}{|x|^2}\right)
  \hbox{\quad as\quad} |x|\to\infty.
  \tag{1.6}\end{equation}

For such a solution to exist, it is necessary, and, as we prove,
sufficient that
\begin{equation}
 \int_{\R^2} f(x)dx=0.
 \tag{1.7}\end{equation}
If (1.5) holds then $f(x)\in L^1(\R^2)$ and (1.7) makes sense.

If one drops the assumption about compactness of the support
of $f(x)$, then one may assume that
\begin{equation}
  |f(x)|=O\left( \frac{1}{|x|^\gamma} \right), \quad \gamma>2,
 \hbox{\quad as\quad} |x|\to\infty, \tag{1.8}\end{equation}
which implies $f\in L^1(\R^2)$.

The necessity of condition (1.7) is obvious if assumptions (1.6) hold.
Indeed, integrate (1.1) to get
\begin{equation}
  \int_{\R^2} f(x)dx=\lim_{r\to\infty}
  \int_{S_r}\frac{\partial u}{\partial r} ds=0,\quad r> R,
  \tag{1.9}\end{equation}
where $S_r:=\{x: |x|=r, x\in \R^2\},$ and
assumptions (1.4) and (1.6) were used.
The sufficiency of (1.7) for the existence of the unique solution to
(1.1), which satisfies conditions (1.6), will be proved below.
The limiting absorption principle (LAP) for equation (1.1) says that
the unique solution of the problem
\begin{equation}
  L u_\varepsilon +iu_\varepsilon=f
  \hbox{\ in\ }\R^2,
  \quad u_\varepsilon\in L^2(\R^2),
  \quad \varepsilon>0,
  \tag{1.10}\end{equation}
converges, in some sense, as $\varepsilon\to 0$,
to the unique solution of (1.1) satisfying (1.6).

Our results can be stated as follows. Let
\begin{equation}
  \|u\|:=\|(1+|x|)|u(x)|\,\|_{L^\infty(\R^2)},
  \quad \||f\||:=\|f\|_{L^2\left(B_R\right)},
  \tag{1.11}\end{equation}
and assume that $u$ satisfies (1.6).

\begin{theorem} 
Let the assumptions (1.2) -- (1.6) hold. Then equation (1.1)
has a solution $u(x)$ such that
\begin{equation}
  \|u\|\leq c\,\||f\||,\quad c=\hbox{\rm const}>0,
  \tag{1.12}\end{equation}
if and only if the assumption (1.7) holds.
Here $c$ is a constant independent on $f$.
This solution is unique, satisfies estimates (1.6) and
can be constructed by the formula
\begin{equation}
  \lim_{\varepsilon\downarrow 0}\|u_\varepsilon(x)-u(x)\|=0,
  \tag{1.13}\end{equation}
where $u_\varepsilon\in L^2(\R^2)$ is the unique solution to
(1.10).
One has
\begin{equation}
  \|u_\varepsilon\|\leq c\,\||f\||,
  \tag{1.14}\end{equation}
where $c=\hbox{\rm const}>0$ is independent of $f$ and
$\varepsilon\in(0,\varepsilon_0)$, and $\varepsilon_0>0$
is an arbitrary small fixed number.
\end{theorem}

Consider the equation
\begin{equation}
  L w_\varepsilon-k^2w_\varepsilon-i\varepsilon w_\varepsilon=f
  \hbox{\ \ in\ \ }\R^2,
  \quad k>0, \quad \varepsilon>0,
  \tag{1.15}\end{equation}
where $k$ is a constant, $\varepsilon\in(0,\varepsilon_0)$.
Let
\begin{equation}
  L w-k^2w=f\hbox{\ \ in\ \ }\R^2, \quad k>0.
  \tag{1.16}\end{equation}
If $k^2=\sigma+i\tau$, $\tau>0$ (one can always define
$k=\sqrt{k^2}$ so
that $\Im\, k\geq 0$), then there is a unique solution to (1.15) and
to (1.16) in $L^2(\R^2)$. If $\tau=0$ and $\sigma< 0$
the same is true.But if $\tau=0$ and $\sigma\geq 0$, one
one has to consider two cases: $k=0$, that is, $\sigma=0$,
and $k>0$, that is, $\sigma>0$.

The first case is the subject of Theorem 1.1.
The second case is treated in Theorem 1.2.
The case when $k\to 0$ along any radial direction,
that is, $k=\kappa e^{i\varphi}$, $\kappa>0$, $0<\varphi<2\pi$,
$\kappa\to 0$, except the ray $k>0,$ is treated as in Theorem
1.1.

The case when $k\to 0$, $k>0$, is treated in Theorem 1.2.
In this case we look for the solution to (1.16) satisfying the
radiation condition:
\begin{equation}
  \lim_{r\to\infty} \int_{S_r}
  \left|\frac{\partial w}{\partial r} -ikw\right|^2ds=0.
  \tag{1.17}\end{equation}
It is well known [5,p.25] that there is at most one solution to
problem (1.16) -- (1.17).

Let us assume (1.2)-(1.6).

\begin{theorem} 
If $k>0$ is fixed and $\epsilon >0$
 then problem (1.16) -- (1-17) has a unique solution
for any $f$ satisfying (1.8). This solution satisfies the inequality
\begin{equation}
  \|w\|^2_{-b}
  :=\int_{\R^2} \frac{|w|^2}{1+|x|^b}dx
  \leq c \int_{\R^2} |f|^2 (1+|x|^b)dx
  :=\|f\|^2_b, \quad b>1,
  \tag{1.18}\end{equation}
and
\begin{equation}
  \|(1+|x|^{1/2})|w|\,\|_{L^\infty(\R^2)}
  \leq c\|f\|_{L^2(B_R)}.
  \tag{1.19}\end{equation}
If $k>0$ is fixed, then problem (1.15) has a unique solution in
$L^2(\R^2)$
for any $f$ satisfying (1.8). This solution converges to $w$:
\begin{equation}
  \|w_\varepsilon-w\|_{-b}\to 0,
  \quad \| \,|w_\varepsilon-w| (1+|x|^{1/2})\|_{L^\infty(\R^2)}\to 0
  \hbox{\ \ as\ \ }\varepsilon\downarrow 0,
  \tag{1.20}\end{equation}
and
\begin{equation}
  \|w_\varepsilon\|_{-b}\leq c\|f\|_b,
  \quad b>1,
  \tag{1.21} \end{equation}
where $c>0$ is independent of $f$ and $\varepsilon$.

If $k>0$ and assumption (1.7) holds in addition to (1.2) -- (1.6),
and $k\to 0$, then
\begin{equation}
  \lim_{k\to 0}\|(w(x,k)-u)|x|^{1/2}\|=0,
  \tag{1.22}\end{equation}
where $u$ is the solution to (1.1), (1.6).

\end{theorem}

\section{Proofs.}
We give a detailed proof of Theorem 1.1 and outline the new points
in the proof of Theorem 1.2, leaving the details to the reader.

\subsection{Proof of Theorem 1.1}
Uniqueness of the solution to (1.1) with decay (1.6) at infinity
follows immediately from the maximum principle and a much
weaker than (1.6) condition $|u(x)|=o(1)$ as $|x|\to\infty$
is sufficient for the uniqueness of the solution to (1.1).

Existence of the solution to (1.1), (1.6) under the assumption (1.7)
will be proved as follows: we prove that the function 
defined as the limit in (1.13), satisfies equation (1.1) and
conditions (1.6), and the limit in (1.13) exists if and only if
assumption (1.7) holds.

Denote
\begin{equation}
  g_0(x,y)=\frac{1}{2\pi}\ln\frac{1}{|x-y|},
  \quad \Delta g_0=-\delta(x-y),
  \tag{2.1}\end{equation}
where $\Delta$ is the Laplacian. If $x\in B'_R$, then (1.1)
can be written as
\begin{equation}
  -\Delta u=f\hbox{\ \ in\ \ }B'_R.
  \tag{2.2} \end{equation}
Let us first assume that a solution to (1.1) and (1.6) exists
and derive (1.7) from this assumption.
By Green's formula one gets, using (1.6),
\begin{equation}
  u(x)=- \int_{B'_R} g_0(x,y) f(y)dy
  -\int_{S_R} \left[ g_0(x,s)u_{N_s}(s)-g_{0N_s}(x,s)u(s)\right]
ds,
  \quad x\in B'_R.
  \tag{2.3}\end{equation}
Here $N_s$ is the normal, at the point $s\in S_R$, to the
circle $S_R:=\{s:|x|=R\}$, pointing into $B'_R$.

If (1.5) holds, then $f=0$ in $B'_R$, and (2.3) yields
\begin{equation}
  u(x)=-\int_{S_R} [g_0(x,s)u_{N_s}(s)-g_{0N_s}(x,s)u(s)] ds.
  \tag{2.4}\end{equation}
If $|x|\to\infty$ and $y$ belongs to a compact set in $\R^2$,
then, uniformly with respect to $y$, one has:
\begin{equation}
  g_0(x,y)=\frac{1}{2\pi}\ln\frac{1}{|x|}
  +O\left(\frac{1}{|x|}\right),
  \quad |x|\to\infty.
  \tag{2.5}\end{equation}

Note that
\begin{equation}
  g_{N_s}(x,s)=-\frac{1}{2\pi}
  \ \frac{\cos(r_{\overrightarrow{xs}},N_s)}{|x-s|}
  = O\left( \frac{1}{|x|}\right),
  \quad |x|\to\infty,
  \quad s\in S_R.
  \tag{2.6}\end{equation}
Therefore (2.4) implies
\begin{equation}
  u(x)=-\frac{1}{2\pi} \ln\frac{1}{|x|}
  \int_{S_R} u_{N_s} ds + O\left( \frac{1}{|x|}\right),
  \quad |x|\to\infty.
  \tag{2.7}\end{equation}
Since we assumed that $u(x)$ solves (1.1) and (1.6), it follows from (2.7)
that
\begin{equation}
  \int_{S_R} u_{N_s} ds=0.
  \tag{2.8}\end{equation}
Note that condition (2.8) would follow from (2.7) under the only
assumption that $u(x)$ is bounded as $|x|\to\infty$.

Therefore we have proved

{\bf Claim 1.} {\it If there is a solution 
to (1.1), (1.6), then (2.8) holds.}

Let us show that (2.8) is equivalent to (1.7).
Indeed, in $B'_R$ equation (2.2) reduces to
\begin{equation}
  -\Delta u=0\hbox{\ \ in\ \ } B_R,
  \tag{2.9}\end{equation}
since $f=0$ in $B'_R$. From (2.9) one gets
\begin{equation}
  \int_{S_R} u_{N_s} ds=\int_{S_r} u_{N_s} ds,
  \quad \forall r>R.
  \tag{2.10}\end{equation}
Thus
\begin{equation}
 \int_{\R^2} f(x)dx=\lim_{r\to\infty} \int_{S_r} u_{N_s}ds=0,
 \tag{2.11}\end{equation}
if (2.8) holds. So (2.8) implies (1.7).

Conversely, if (1.7) holds, then (2.8)
follows from (2.11),  and (1.6) holds by formula (2.4) for the solution
to (1.1).

The main difficulty of the proof is to establish existence of the
solution to (1.1) and representation formula (2.4) for this solution
which allows one to prove (2.7) and to derive (1.6).

Let us now assume that (1.7) holds and derive the existence
of the solution to (1.1), (1.6).
Denote by $g_\varepsilon(x,y)$ the Green function
(resolvent kernel of $-\Delta+i\varepsilon$):
\begin{equation}
  (-\Delta+i\varepsilon)g_\varepsilon(x,y)
  =\delta(x-y)\hbox{\ \ in\ \ }\R^2,
  \tag{2.12} \end{equation}
where $g_\varepsilon(\cdot,y)\in L^2(\R^2)$.
It is well known (see e.g. [8, p.20-21]) that
\begin{equation}
  g_\varepsilon(x,y)=\alpha(\varepsilon)+g_0(x,y)
    +O(\varepsilon^2\ln\varepsilon)
  \hbox{\ \ as\ \ }\varepsilon\to 0,
  \tag{2.13}\end{equation}
where $g_0(x,y)$ is defined in (2.1), 
$O(\varepsilon^2\ln\varepsilon)$ is uniform in the region
$0<c_1\leq|x-y|\leq c_2<\infty$, 
and 
$\alpha(\varepsilon)=O(\ln\varepsilon)$
as $\varepsilon\to 0$, $\varepsilon>0$.

Let $u_\varepsilon\in L^2(\R^2)$ be the (unique) solution to (1.10).
Such a solution exists since $L$ is selfadjoint in $L^2(\R^2)$.

Using (1.5) and the formula analogous to (2.4), one gets:
\begin{equation}
  u_\varepsilon(x)=-\int_{S_R} [g_\varepsilon(x,s)u_{\varepsilon N_s}
   -g_{\varepsilon N_s}(x,s) u_\varepsilon(s)]ds.
   \tag{2.14} \end{equation}

Assume for a moment that
\begin{equation}
 \sup_{0<\varepsilon\leq\varepsilon_0} \|u_\varepsilon\|\leq c\| |f\| |,
 \tag{2.15} \end{equation}
where the norms are defined in (1.11), the constant $c>0$
is independent of $f$, and $\varep\in(0,\varep_0)$.
Then $u_\varep(x)$ is bounded on compact subsets of $\R^2$.
Therefore, by elliptic regularity,
$u_\varep\in H^2_{loc}(\R^2)$ (the Sobolev space)
if $a_{ij}\in C^{0,1}(\R^2)$ (and $u_\varep\in H^1_{loc}(\R^2)$ if
$a_{ij}\in L^\infty(\R^2)$).

This implies that $u_\varep\rightharpoonup u_0$ 
in $H^2_{loc}(\R^2)$, where
$\rightharpoonup$ stands for the weak convergence. 
Using the known estimate
\begin{equation}
 \|u\|_{H^2(D_1)}\leq c \left(\| Lu\|_{L^2(D_2)}+\|u\|_{L^2(D_2)}\right),
 \tag{2.16}\end{equation}
where $D_1$ is a strictly inner subdomain of a bounded domain $D_2$,
one concludes that $u_\varep\to u_0$ in $H^2_{loc}(\R^2)$,
where $\to$ stands for the strong convergence.
Thus one can pass to the limit in the equation (1.10) and conclude
that $u_0$ solves (1.1). Also one can pass to the limit in (2.14).
By the embedding theorems convergence in $H^2_{loc}(\R^2)$ implies
\begin{equation}
  u_\varep\to u_0\hbox{\ \ in\ \ }H^{3/2}(S_R),
  \quad u_{\varep N_s}\to u_{0N_s} \hbox{\ \ in\ \ }H^{1/2}(S_R).
  \tag{2.17} \end{equation}
Passing to the limit $\varep\downarrow 0$ in (2.14) and using (2.13),
one gets as a necessary condition for the existence of the limit
condition (2.8) for $u_{0N_s}$. 
Also one gets the representation formula for $u_0$:
\begin{equation}
  u_0(x)=-\int_{S_R}
  [g_0(x,s)u_{0N_s}(s)-g_{0N_s}(x,s)u_0(s)] ds,
  \quad x\in B'_R,
  \tag{2.18} \end{equation}
\begin{equation}
  \int _{S_R} u_{0N_s}(s) ds=0.
  \tag{2.19} \end{equation}
From (2.18) and (2.19) one concludes, as above, that $u_0(x)$ satisfies
conditions (1.6). We have already proved that there is at most one
solution to (1.1), (1.6). The constructed function $u_0(x)$ is this
solution and $\|u_0(x)\|<\infty$.
To complete our argument, estimate (2.15) has to be proved.

{\bf Claim 2.} {\it Estimate (2.15) holds.}

To prove this claim assume the contrary. Then there exists a sequence
$\varepsilon_n\downarrow 0$ and $f_n$, $\||f_n\||=1$, such that
\begin{equation}
  \|u_{\varep_n}\| \geq n.
  \tag{2.20}\end{equation}

Let $u_{\varep_n}:=u_n$, $v_n:=\frac{u_n}{\|u_n\|}$. Then
\begin{equation}
  L v_n+i\varep v_n=\frac{f_n}{\|u_n\|}:=h_n,
  \tag{2.21}\end{equation}
\begin{equation}
 \|v_n\|=1,\quad \||h_n\||\to 0 \hbox{\ \ as\ \ } n\to\infty.
 \tag{2.22} \end{equation}
Repeating the above argument, one gets a subsequence, which is
denoted $v_n$ again, such that
\begin{equation}
  v_n\to v\hbox{\ \ in\ \ } H^2_{loc} (\R^2),
  \tag{2.23}\end{equation}
and
\begin{equation}
  Lv=0,
  \tag{2.24} \end{equation}
since $h_n\to 0$. Moreover, $v$ satisfies (1.6) and
\begin{equation}
  \|v\|<\infty.
  \tag{2.25} \end{equation}
By the  already proved uniqueness theorem it follows that
$v=0$. To get a contradiction with the first relation (2.22),
let us prove that
\begin{equation}
  \lim_{n\to \infty} \|v_n\|=0.
  \tag{2.26} \end{equation}
If (2.26) is verified the proof of Claim 2 is complete.
However, on compacts $v_n\to 0$ pointwise, as we have proved, and at
infinity one has
\begin{equation}
  (1+|x|)|v_n(x)|\to 0 \hbox{\ \ as\ \ } |x|\to\infty,
  \tag{2.27} \end{equation}
as follows from the representation formula analogous to (2.14) for
$v_n$, and from the condition
\begin{equation}
  \lim_{n\to\infty} \alpha(\varep_n) \int_{S_R} v_{nN} N_s ds=0,
  \tag{2.28} \end{equation}
which follows from our argument used above.

Claim 2 is proved. \qed

Theorem 1.1 is proved. \qed

\subsection{Proof of Theorem 1.2.}

Proof of Theorem 1.2 follows the pattern of the proof of Theorem 1.
Therefore, we indicate only the main new points (enumerated below)
and leave the details to the reader.

\begin{description}
\item[1.]
If $k>0$ is fixed, then equation (1.16) has at most one
solution satisfying the radiation condition (1.7).

This is an immediate consequence of the Rellich-type lemma
\cite[page~25]{R1} and the unique continuation theorem for
elliptic equations (this theorem requires
$a_{ij}\in C^{0,1}$).

Indeed, if $(L-k^2)w=0$ then (1.5) implies
\begin{equation}
  (-\Delta-k^2)w=0\hbox{\ \ in\ \ } B'_R,\quad k>0.
  \tag{2.29} \end{equation}
From the real-valuedness of $a(x)$ and $f(x)$ one gets:
\begin{equation}
  0=\lim_{r\to\infty} \int_{B_r}
  [\nabla\cdot(a\nabla w)\overline{w}-w\nabla\cdot(a\nabla\overline{w})] ds
  =\lim_{r\to 0} \int_{S_r} (w_N\overline{w}-w\overline{w}_N) ds.
  \tag{2.30} \end{equation}
From (2.30) and (1.17) one concludes:
\begin{equation}
  \lim_{r\to\infty} \int_{S_r} (|w_r|^2 +k^2|w|^2)ds=0.
  \tag{2.31}\end{equation}
From (2.31) and (2.29) it follows \cite[page~25]{R1}
that
\begin{equation}
  w=0\hbox{\ \ in\ \ } B'_R.
  \tag{2.32}\end{equation}
By the unique continuation theorem, (2.32) implies that the solution to
\begin{equation}
  (L-k^2)w=0\hbox{\ \ in\ \ } \R^2
  \tag{2.33}\end{equation}
vanishes identically, $w\equiv 0$. \qed

\item[2.]
Representation formula for the solution to (1.16) is analogous to (2.4):
\begin{equation}
  w(x,k)=-\int_{S_R} [G(x,s,k)u_{N_s}(s,k)-G_{N_s}(x,s,k)w(s,k)]ds,
  \tag{2.34}\end{equation}
where
\begin{equation}
  G(x,y,k):=\frac{i}{4} H^1_0(k|x-y|),
  \quad (\nabla^2+k^2)G=-\delta(x,y)\hbox{\ \ in\ \ } \R^2.
  \tag{2.35}\end{equation}

From (2.34) it follows that
\begin{equation}
  |w(x,k)| + |\nabla w(x,k)| = O\left( \frac{1}{|x|^{1/2}} \right)
  \hbox{\ \ as\ \ } |x|\to\infty.
  \tag{2.36} \end{equation}
This is the reason for the weight $(1+|x|^{1/2})$ in (1.19)
and for the exponent $b>1$ in (2.18):
\begin{equation}
  |w|^2=O\left(\frac{1}{|x|}\right),
  \quad\frac{|w|^2}{1+|x|^b} \in L^1(\R^2)
  \hbox{\ \ if\ \ }b>1.
  \tag{2.37} \end{equation}
The proof of the limiting absorption principle goes essentially as in the
proof of Theorem 1 and is left to the reader.

\item[3.]
If $k^2$ approaches zero along any ray $k=\varep e^{i\varphi}$,
$0<\varphi<2\pi$, the proof of the LAP is essentially the proof given
in Theorem 1 (where $\varphi=\frac{\pi}{2}$ is used).

If $\varphi=0$ (or $\varphi=2\pi$, which is the same for our
argument) then there is a new point explained in items 1 and 2 above.
This new point consists of two parts:

\begin{description}
\item[a)] uniqueness class for the solutions of equation (1.16)
is defined by the radiation condition (1.17),
\item[b)] the asymptotics at infinity of the solution is given by (36)
rather than (1.6).
\end{description}

\end{description}

\begin{remark} 
One could drop assumption (1.5) and use (1.8)
(or the boundedness of the norm $\|f\|_{2,s}<\infty$,
where $\|f\|^2_{2,s}:=\int_{\R^2}|f|^2(1+|x|^2)^s dx $).
Such assumptions would lead to additional technical 
difficulties,
namely, in the estimates of the integral
\begin{equation}
  \int_{B'_R}g_\varep(x,y)f(y)dy
  \tag{2.38}\end{equation}
in some integral norms with weights. These difficulties are not
essential. 

In fact, in the basic estimate (2.15) one can use as 
the norm of $f$ any norm the boundedness of which
allows one to prove that function (2.38) is
bounded in the norm  $||\cdot||_{-b}$ (1.18) with $b>1$,
( or in the norm (1.11) for $u$)
uniformly with respect to $\epsilon\in (0,\epsilon_0)$,
and the scheme developed in section 1 goes through.
 
\end{remark}

\begin{remark} 
The assumption $f=\overline{f}$ is not important and can be dropped.
If $f=f_1+if_2$, where $f_1$ and $f_2$ are real-valued,
then the solution to (1.1) can be written as $u=u_1+iu_2$, where
$u_1$ and $u_2$ are some real-valued functions.
Here the real-valuedness of $a_{ij}$ is used.
Therefore equation (1.1) can be studied for real-valued $f$.

Without loss of generality one can also study equation (1.16)
assuming $f(x)$ is real-valued. In this case $u_1$ and $u_2$
are not real-valued since the radiation condition (1.17) is not
self-adjoint. However, the real-valuedness of  $u_1$ and $u_2$ is
not important for this argument. What is important is the
uniqueness of the solutions to homogeneous versions of equations
(1.1) and (1.16) in the class of functions satisfying conditions
(1.6) and (1.17) respectively.
\end{remark}

\vfill\pagebreak

\end{document}